\documentclass[final,5p,times,twocolumn]{elsarticle}
\usepackage{graphicx}
\usepackage{graphicx}% Include figure files
\usepackage{dcolumn}% Align table columns on decimal point
\usepackage{bm}% bold math
\usepackage[T1]{fontenc}
\usepackage[usenames]{color}

\large\newcommand{\D}{\displaystyle}

\newcommand{\be}{\begin{equation}}
\newcommand{\bel}{\begin{equation}\label}
\newcommand{\ee}{\end{equation}}
\def\dg{\dagger}
\newcommand{\barl}{\begin{eqnarray}\label}
\newcommand{\ear}{\end{eqnarray}}

\usepackage{amssymb}
\usepackage{amsmath}

\journal{Physica B}

\begin{document}
\def\bw{\begin{widetext}}
\def\ew{\end{widetext}}
\def\dg{\dagger}
\def\d{\dj{}}
\def\D{\DJ{}}
\newcommand{\al }{$\alpha$--helix}
\begin{frontmatter}

%% Title, authors and addresses

%% use the tnoteref command within \title for footnotes;
%% use the tnotetext command for the associated footnote;
%% use the fnref command within \author or \address for footnotes;
%% use the fntext command for the associated footnote;
%% use the corref command within \author for corresponding author footnotes;
%% use the cortext command for the associated footnote;
%% use the ead command for the email address,
%% and the form \ead[url] for the home page:dvk@df.uns.ac.rs
%%
%% \title{Title\tnoteref{label1}}
%% \tnotetext[label1]{}
%% \ead[url]{home page}
%% \fntext[label2]{}
\cortext[cor1]{Corresponding author}
%% \address{Address\fnref{label3}}
%% \fntext[label3]{}

\title{On the vibron nature in the system of two parallel macromolecular chains: the influence of interchain coupling}
%% use optional labels to link authors explicitly to addresses:
\author[innv]{Dalibor \v Cevizovi\'c \corref{cor1}}
\ead{cevizd@vinca.rs}
\author[innv,ccfqcn,misis]{Zoran Ivi\'c}
\author[innv]{Slobodanka Galovi\'c}
\author[ispms]{Alexander Reshetnyak}
\author[jinr,dubnauniversity]{Alexei Chizhov} 
\address[innv]{University of Belgrade, "Vin\v ca" Institute of Nuclear sciences, Laboratory for Theoretical and Condensed Matter Physics, P.O. BOX 522, 11001, Belgrade, Serbia}
\address[ccfqcn]{Crete Center for Quantum Complexity and Nanotechnology, Department of Physics, University of Crete, P. O. Box 2208, 71003 Heraklion, Greece}
\address[misis]{National University of Science and Technology MISiS, Leninsky prosp. 4, Moscow, 119049, Russia}
\address[ispms]{Institute of Strength Physics and Materials Science SB RAS, Tomsk, 634055, Russia}
\address[jinr]{Joint Institute for Nuclear Research, Bogoliubov Laboratory of Theoretical Physics, Dubna, 141980, Russia}
\address[dubnauniversity]{Dubna International University, Dubna, 141980, Russia}

\begin{abstract}
We studied the properties of the intramolecular vibrational excitation (vibron) at finite temperature in a system which consists of two parallel macromolecular chains. It was assumed that vibron interacts exclusively with dispersionless optical phonons and the whole system is considered to be in thermal equilibrium. Particular attention has been paid to the examination of the impact of the temperature and strength of the interchain coupling on the \emph{small polaron} crossover. For that purpose we employed partial dressing method which enables the study of the degree of the phonon dressing of the vibron excitations in a wide area of system parameter space. We found that in the non--adiabatic regime the degree of dressing as a function of coupling constant continuously increases reflecting the smooth transition of the slightly dressed, practically free vibron, to a heavily  dressed one: small polaron. As "adiabaticity" rises this transition becomes increasingly steeper, and finally, in the adiabatic limit, a discontinuous "jump" of the degree of dressing is observed. The interchain coupling manifests itself through the increase of the effective adiabatic parameter of the system.
\end{abstract}

\begin{keyword}
%% keywords here, in the form: keyword \sep keyword
vibron\sep molecular chain\sep polaron\sep interchain coupling
%% MSC codes here, in the form: \MSC code \sep code
%% or \MSC[2008] code \sep code (2000 is the default)

\end{keyword}

\end{frontmatter}

%%
%% Start line numbering here if you want
%%
% \linenumbers
%% main text

\section{Introduction}

Coherent long--range transfer of the intramolecular (IMO) vibrational energy along the protein macromolecules plays the crucial role in the conversion of the energy released in the hydrolysis of the adenosine triphosphate (ATP) into the mechanical work \cite{DavydovJTB,DavydovPS,DavydovPD}. Elucidation of the high efficiency of these transfer processes has been the open question for years. Numerous different models have been proposed \cite{DavydovJTB,DavydovPS,DavydovPD,AK,BI}, however, the underlying microscopic mechanisms governing IMO transport processes are not fully understood so far. Appealing model of the IMO transfer has been proposed by Davydov and his co-workers \cite{DavydovJTB,DavydovPS,DavydovPD}, who suggested that the long range IMO vibrational energy transfer in biological macromolecules is achieved by means of the self--trapping (ST) mechanism: the energy released in the hydrolysis of the ATP may be resonantly absorbed by the protein molecule as a quanta of the CO stretching oscillations (Amide--I quanta) whose transfer along the molecule spine is facilitated by the dipole--dipole coupling of neighbouring peptide groups. According to the Davydov theory the IMO vibrational energy transport is provided by the strong interaction of the Amide--I quanta with acoustic phonons which lead to a local distortion of a molecule and polaron formation. Quasi--one dimensional (Q1D) structure of DNA, $\alpha$--helix proteins and some other biological macromolecules plays the crucial role in stability of the excitation transfer since a polaron in Q1D media acquires a soliton form provided that the dipole--dipole interaction (i.e. excitation bandwidth) highly exceeds both maximal phonon and lattice deformation energies: i.e. the adiabatic and strong coupling conditions are required for the soliton formation \cite{Rashba}.

The main support in favor of the vibron ST in protein macromolecular chains (MC) has been found in the experiments of the infra--red (IR) absorption spectra of the crystalline acetanillide (ACN), which may be regarded as a "model protein" molecule \cite{AK} due to the its striking structural similarities with the natural hydrogen--bonded polypeptide chains. Nevertheless, the values of the basic energy parameters of ACN lie beyond the applicability of continuum adiabatic theory and soliton model is inappropriate for the the description of the self--trapping in these media \cite{AK,CevizovicCPL2008,PouthierJCP132}. Improved theory was presented by Alexander and Krumhansl who interpreted the IR absorption spectra of ACN \cite{AK} in terms of the one--dimensional small polaron (SP) model. Subsequent theoretical investigation \cite{AK,BI,CevizovicCPL2008,AlvinScottPR217,CevizovicCPL2009} has shown that the SP theory can explain, in principle, some of the open questions in the problem of transport dynamics of IMO. Nevertheless, even so improved Davydov theory, being entirely based on pure 1D models, can not be fully reliable for the objective description of the excitation transfer in the realistic Q1D structures which are composed of the two (ACN, DNA), three ($\alpha$--helix) or more coupled MCs (some conjugated polymers and Q1D conductors). In particular, results of some theoretical studies \cite{Emin,PertzchSSC37,IvicCP426,IvicCSF73} have shown that even tiny interchain coupling can substantially affect the character of the ST states. This especially concerns the adiabatic limit where, in the highly anisotropic solids composed of the large number parallel MC, there exists the threshold value, of the order of $10^{-2}$, of the ratio of the interchain over the intrachain transfer integrals above which the large polaron (soliton) becomes unstable. Below these critical coupling soliton exhibits quite interesting features depending the number of chains and their particular geometrical arrangement \cite{IvicCP426,IvicCSF73}.

Properties of small--polarons in a systems consisting of the finite number of coupled molecular chains have not been the subject of comprehensive theoretical analysis so far. In particular, Pouthier and Falvo \cite{FalvoJCP123a,FalvoJCP123b} have studied the SP states in $\alpha$--hekix, within the conventional SP theories employing Lang--Firsov unitary transformation method. However, such an approach apply in strict non--adiabatic and strong coupling regime, cannot give a reliable description of small polarons in realistic substances in which the values of the physical parameters span a quite wide area of system parameter space interpolating between non--adiabatic and adiabatic as well as between strong and weak coupling regimes \cite{PouthierJCP132,NevskayaBiopolym15,ScottPRA26,ScottPR217}. Under these circumstances, vibrons are only partially dressed and their study requires theoretical approaches that go beyond the standard SP theories. In \cite{HammPRB73} Edler and Hamm presented a numerically exact solution of the Holstein polaron Hamiltonian in 1D and 3D using the concept that has been introduced by Trugman and co-workers \cite{TrugmanPRB81}. Their primary interest has been to elucidate the effects of dimensionality on the peculiar IR spectra of ACN. For that reason, the impact of the values of system parameters and temperature on the nature of the eigenstates of the model were not given in a transparent and systematic way.

In this paper we study the influence of the temperature and the interchain coupling on the properties of the small polarons in media composed of the two coupled infinite parallel molecular chains. For that purpose we employ variational treatment based on the modified Lang--Firsov transformation \cite{BI,PouthierPRB79,FujinashiJPCB}.

The work is organized as follows. In Section 2, we formulate the theoretical models for two cases of parallel MC being unshifted and shifted from each other for which we suggest the respective Hamiltonians, and calculate an optimal value for the variational parameter. The results of our study and concluding comments on their physical interpretation are given in Section 3.

\section{Theoretical model}

As a theoretical framework of our study, we use the Holstein molecular crystal model modified to account for the excitation transfer between the chains: 

\begin{align}
H&=
E_0\sum_{n,j}{a^{\dg}_{j,n}a_{j,n}}-J\sum_{n,j}{a^{\dg}_{j,n}(a_{j,n-1}+a_{j,n+1})}+\sum_{q,j}{\hbar\omega_qb^{\dg}_{j,q}b_{j,q}}+\nonumber\\&
+\frac{1}{\sqrt{N}}\sum_{n,q,j}{F_q\mathrm{e}^{iqnR_0}a^{\dg}_{j,n}a_{j,n}(b_{j,q}+b^{\dg}_{j,-q})}+H_L
\end{align}

\noindent where $H_L$ is the term describing the interchain dipole--dipole interaction which depends on the macromolecule geometry. Here, two different cases will be considered. The first one corresponds to a situation in which the structural units (molecular groups) of both strands are positioned exactly opposite one to another, Fig.1. (upper panel). The second case refers to shifted chains in which the units in one chain are shifted half of the interchain lattice constant relative to those in the other one, Fig.1. (lower panel). In the case of unshifted MCs interchain dipole--dipole interaction term becomes:

$$
H_L=L\sum_n{\left(a^{\dg}_{1,n}a_{2,n}+a^{\dg}_{2,n}a_{1,n}\right)}
$$

\noindent while in the case of the shifted MCs it has the form:

$$
H_L=L\sum_n{\left\{a^{\dg}_{1,n}a_{2,n}+a^{\dg}_{1,n}a_{2,n+1}+a^{\dg}_{2,n}a_{1,n-1}+a^{\dg}_{2,n}a_{1,n}\right\}}
$$

\begin{figure}[h]
    \begin{center}
\includegraphics[height=8.0 cm]{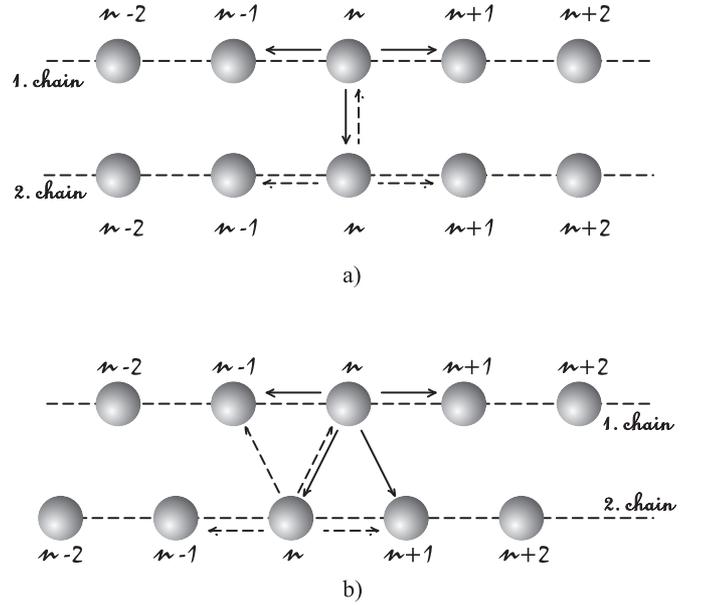}
        \caption{Grphical representation of the considered model for the unshifted (upper panel) and shifted (lower panel) MCs. The possible leaps of the vibron excitation from $n$--th structure element are presented by arrows, in the approximation of nearest neighbours.}\label{f1}
    \end{center}
%    \textit{}
\end{figure}

Index $n$ counts the molecular group units within the $j$--th chain ($j=1,2$); $E_0$ is the on--site excitation energy of the vibron, $a^{\dg}_{j,n}$ ($a_{j,n}$) are vibron creation (annihilation) operators, $J$ and $L$ are the nearest neighbour intra--chain and inter--chain dipole--dipole interaction parameters, respectively; $b^{\dg}_{j,q}$ ($b_{j,q}$) are creation (annihilation) operators of the $q$--th phonon mode at $j$--th chain with the frequency $\omega_q$; finally, $F_q$ is the vibron--phonon interaction parameter. Mechanical subsystems (phonons) in both chains will be considered to be independent of each other. We restricted ourselves to the case of the vibron interaction with non--dispersive optic phonon modes since the experiments of the incoherent neutron scattering have indicated that only those modes are involved in the processes of the vibron ST in molecular structures such as ACN and $\alpha$--helix \cite{BarthesJML}. Accordingly we took $F_q=F_0$ and $\omega_q=\omega_0$.

In order to describe the properties of vibron ST in such structures, we introduce adiabatic parameter $B=\frac{2J}{\hbar\omega_0}$, the coupling constant $S=\frac{E_B}{\hbar\omega_0}$ and interchain coupling strength $g=\frac{L}{J}$ \cite{BI,Emin,IvicCP426,IvicCSF73}. Here $E_B=\frac{1}{N}\sum_q{\frac{\left|F_q\right|^2}{\hbar\omega_q}}\equiv \frac{F_0^2}{\hbar\omega_0}$ is so called SP binding energy. First of these two parameters determines the character of the lattice deformation engaged in the polaron formation, while the second one determines the polaron spatial size and the degree of quasiparticle dressing. 

Contrary to the problem of the single excitation interacting with the dispersionless optical phonons at zero temperature where numerically exact solutions were found recently \cite{TrugmanPRB81}, the finite temperature treatment is much more complicated and exact solutions, both analytical or numerical, are yet unknown. For this reason we base our study on an approximate, variational mean field approach, which has been used for years as a theoretical framework of the understanding of the various phenomena in the strongly coupled exciton--phonon systems \cite{BI,PouthierPRB79,YarkonyCP,EminPRL,SilbeyJCP128}. At zero temperature this approach reduces to the variational theories of Toyozawa--Merriefild--Emin (TME) \cite{Emin,ToyozawaPTP,MerrifieldJCP40} whose predictions for 3D systems are consistent with the numerical ones \cite{TrugmanPRB81}. This especially concerns the sudden SP crossover when the excitation effective mass in the adiabatic regime discontinuously jumps for a few orders of magnitude when the coupling constant approaches some critical value being different for each adiabatic parameter. Until recently, such singularities in the properties of quasi--particles have been considered as an unphysical artifact of the oversimplified variational method. Nevertheless, the appearance of this "phase transition" has been undoubtedly confirmed by Trugman and his co--workers \cite{TrugmanPRB81} who found that in 3D systems in the non--adiabatic regime the excitation effective mass continuously increases as a function of the coupling constant. In the adiabatic limit this dependence becomes discontinuous when the coupling constant approaches critical value. On the other hand in 1D media the excitation effective mass diverges for any finite coupling constant when phonon frequency tends to zero i.e. in the extreme adiabatic limit. 

Further arguments in favour of the use of this simple method in the present context in spite of its limited validity in 1D systems we found in the study \cite{IvicPLA172} where the problem of the self--trapping in the two--level system interacting with one, two, and three--dimensional phonons have been considered. It was found that so called dimensional crossover may appear at finite temperature: in particular, one--dimensional system at finite temperatures effectively behaves as two-- or three-- dimensional system.

We now pass to the dressed vibron picture, which can be realized by means of the modified Lang--Firsov unitary transformation (MLFUT) \cite{YarkonyCP,CevizovicPRE,CevizovicCPB}. For that purpose we employ the unitary operator $U=U_1\cdot U_2$ where

$$
U_j=\exp{\left\{-\frac{1}{\sqrt{N}}\sum_{q,n}f_{j,q}\mathrm{e}^{-iqnR_0}a^{\dg}_{j,n}a_{j,n}(b_{j,-q}-b^{\dg}_{j,q})\right\}}.
$$

\noindent 
Here $f_{j,-q}=f^*_{j,q}$ are variational parameter(s) to be determined by means of the simple mean--field procedure \cite{BI,CevizovicCPL2008,YarkonyCP,CevizovicPRE,CevizovicCPB}. In accordance with the assumption that the considered chains are identical and that the phonon subsystems in different chains are independent, one may take $f_{1,q}=f_{2,q}=f_q$. 

In order to facilitate an analytic treatment, we assume the equal dressing fraction for all phonon modes introducing the new variational parameter--dressing fraction ($\delta$) taking \cite{BI,PouthierPRB79,CevizovicCPB} $f_q=\delta\frac{F^*_q}{\hbar\omega_q}$. Here $0\leq\delta\leq 1$ measures the degree of the narrowing of the vibron band. In such a way its value determines whether the vibron motion takes place in the coherent fashion (low values of $\delta$) by means of the band mechanism, or it is achieved incoherently by the random jumps between neighbouring sites (high value of $\delta$) \cite{BI,PouthierPRB79,CevizovicCPB}. Such a single parameter method is not flexible enough to provide an accurate description of dressed states in the whole parameter space. This especially concerns the adiabatic regime where the assumption of equal dressing for all phonon modes produces the underestimates  of the critical parameters. Nevertheless, this is not the problem in the case of vibrons in ACN and $\alpha$--helix where values of the system parameters lie in the non--adiabatic and weak to intermediate coupling regime, where the present method gives equivalent results to those obtained by means the improved variational treatments \cite{AK,CevizovicPRE} and numerical calculations \cite{TrugmanPRB81}.

In order to account for the influence of thermal fluctuations on the properties of self--trapped vibron states, we applied a simple mean--field procedure, by averaging of the transformed Hamiltonian over the phonon subsystems \cite{YarkonyCP,CevizovicCPB,LF}. In the first step, we decompose the transformed Hamiltonian $\tilde{H}=U^{\dg}HU$ in a three components by adding and subtracting its thermal average $\langle\tilde{H}\rangle_{ph}$. In this way we obtain $\tilde H=\tilde{H}_0+\tilde{H}_{rest}$, where $\tilde{H}_0=\tilde{H}_{vib}+\tilde{H}_{ph}$ stands for the effective mean field Hamiltonian $\tilde{H}_{vib}=\left\langle \tilde{H}-\tilde{H}_{ph}\right\rangle_{ph}$, and $\tilde{H}_{rest}=\tilde{H}-\tilde{H}_{ph}-\left\langle \tilde{H}-\tilde{H}_{ph}\right\rangle_{ph}$. The symbol $\left\langle\;\right\rangle_{ph}$ denotes the averaging over new--phonon ensembles which are in thermal equilibrium state at temperature $T$. Performing the Fourier transform of the dressed vibron operators ($a_{j,k}=1/\sqrt{N}\sum_na_{j,n}e^{iknR_0}$ ), effective mean field Hamiltonian attains the simple form:

\begin{equation}\label{H2b}
\tilde{H}_{vib}=\sum_{j,k}{E(k)a^{\dg}_{j,k}a_{j,k}}+\sum_k{\left(\lambda_ka^{\dg}_{1,k}a_{2,k}+\lambda^*_ka^{\dg}_{2,k}a_{1,k}\right)}.
\end{equation}

\noindent In the case of unshifted chains $\lambda_k=L\mathrm{e}^{-\delta^2{W}(\tau)}$, (and it is independent of $k$), while for shifted case we have $\lambda_k=L\mathrm{e}^{-\delta^2{W}(\tau)}\left(1+\mathrm{e}^{ikR_0}\right)$. In both cases the vibron band energy is $$E(k)=E_0+\delta(\delta-2)E_B-2J\mathrm{e}^{-\delta^2W(\tau)}\cos kR_0.$$ The narrowing of the intra-- and inter--chain dipole--dipole interaction is characterized by the same factor: $W(\tau)=S{\cal G}(\tau)$. Temperature scaling factor is explicitly ${\cal G}(\tau)=\coth \left(1/2\tau\right)$, and $\tau$ is normalized system temperature $\tau=k_BT/\hbar\omega_0$. Hamiltonian (\ref{H2b}) can be easily diagonalized by means the transformation:

\begin{equation}\label{Tr2}
\alpha_{\mu,k}=\xi^*_k a_{1,k}+{\mu}\xi_k a_{2,k},\;\; \mu=\pm 1.
\end{equation}

For unstagered chains $\xi_k=\frac{1}{\sqrt{2}}$, while for the shifted ones $\xi_k=\frac{\mathrm{e}^{ikR_0/4}}{\sqrt{2}}$. It turns out that vibron spectrum splits in two branches, symmetric and antisymmetric ones, in accordance with the relation

\begin{equation}\label{Hv1}
\tilde{H}_{vib}=\sum_{k,\mu}{E_{\mu}(k)\alpha^{\dg}_{\mu,k}\alpha_{\mu,k}}, \; E_{\mu}(k)=E(k)+\mu \sqrt{|\lambda_k|^2}.
\end{equation}

\section{Results and discussion}

We now determine the optimal value of the dressing fraction in accordance with the Bogoliubov theorem, minimizing so called the upper bound of free energy of the system (trial free energy), which corresponds to the system of the free quasiparticles: dressed vibron and new phonons in the distorted chains:

$$\mathcal{F}_B=-k_BT\mathrm{lnTr}\left\{\mathrm{e}^{-H_0/k_BT}\right\}+\left\langle H_{rest}\right\rangle_{H_0}.$$ 

Note that the second term in this expression is identically equal to zero, while the phonon part of free energy does not depend on $\delta$. Thus for the practical calculation we may consider only the dressed--vibron subsystem. Moreover, having in mind that our analysis concerns the Amide--I excitation in ACN and similar MCs where the excitation energy is $\sim E_0=1665 cm^{-1}$ which corresponds to temperatures of the order of $\approx 2400$ K, we need to consider only the single vibron case so that the basis state is given by  
$\alpha_{\mu,k}^{\dg}|0\rangle_{vib}\prod_j|m_j\rangle$. Accordingly, the effective trial free energy reads

\begin{equation}\label{FB}
\mathcal{F}_B=-k_BT\mathrm{ln}\sum_{\mu,k}{\mathrm{e}^{-E_{\mu}(k)/k_BT}}.
\end{equation}

Requiring $\partial \mathcal{F}_B/\partial \delta$=0 we obtain the transcendent equation for the dressing fraction as a function of the system parameters and temperature $\delta=\delta(S,B,g,T)$ whose solutions we discus in dependence of the values of the adiabatic parameter $B$, coupling constant $S$, and the strength of the interchain coupling $g$.

\begin{figure}[t]
%    \begin{center}
\includegraphics[height=5 cm]{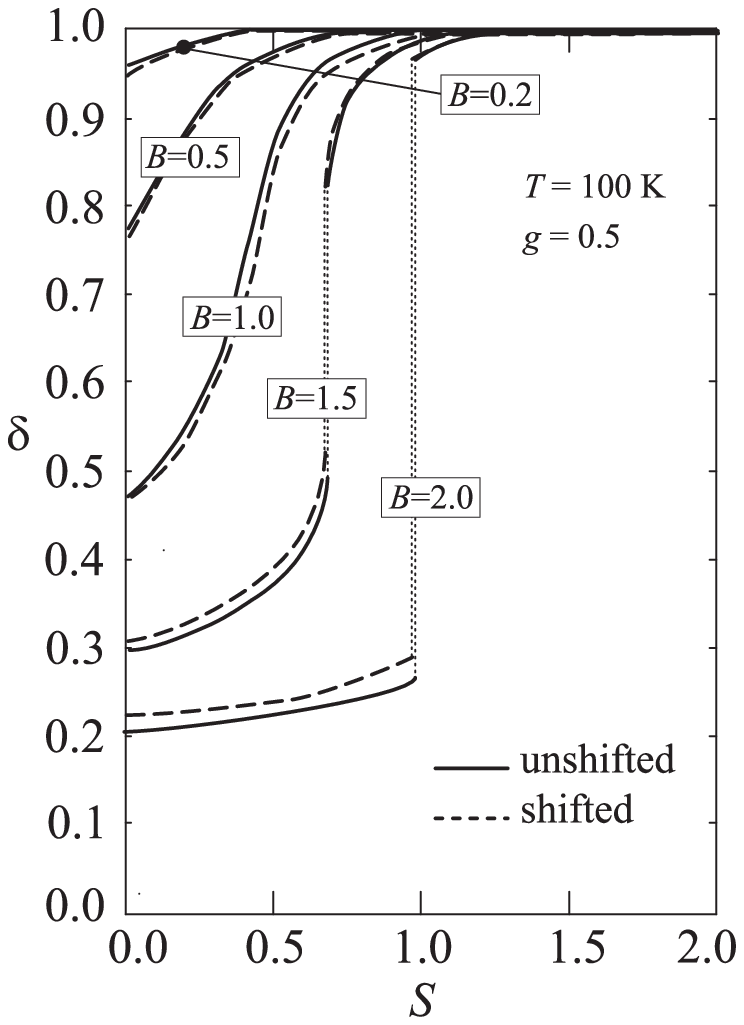}
a
\includegraphics[height=5 cm]{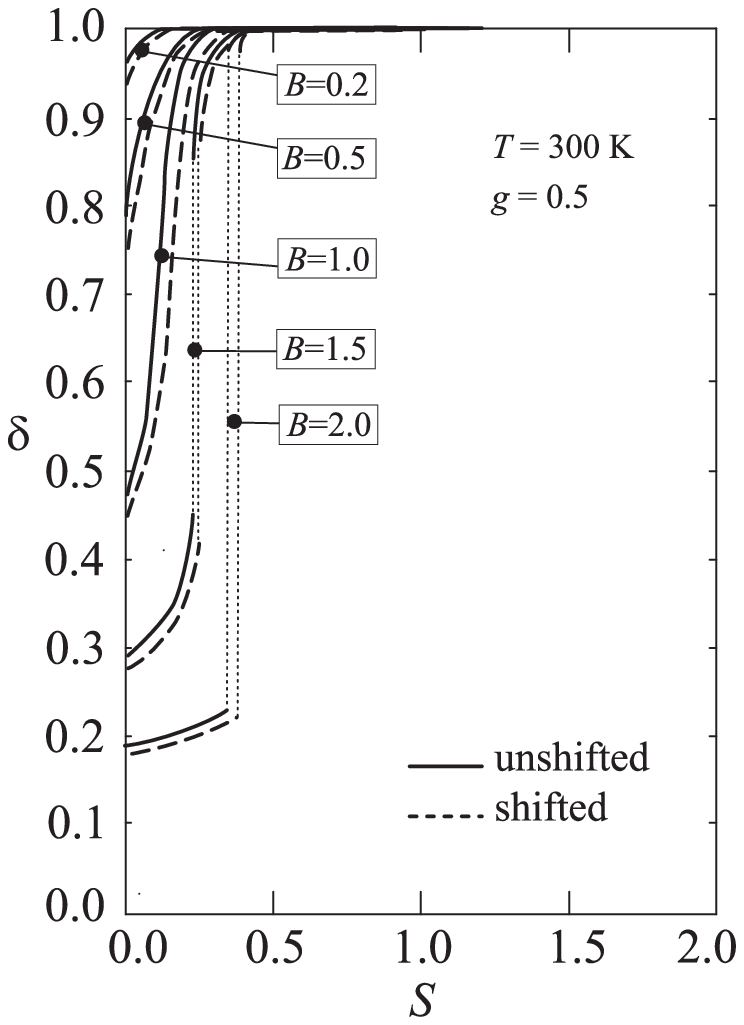}

\includegraphics[height=5 cm]{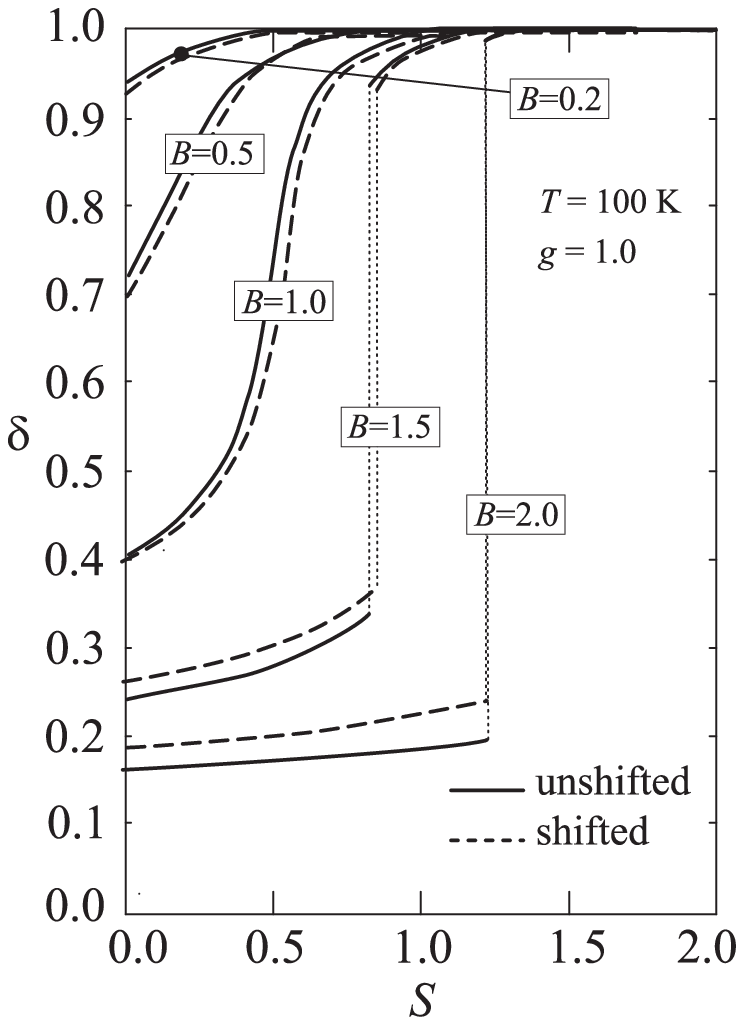}
b
\includegraphics[height=5 cm]{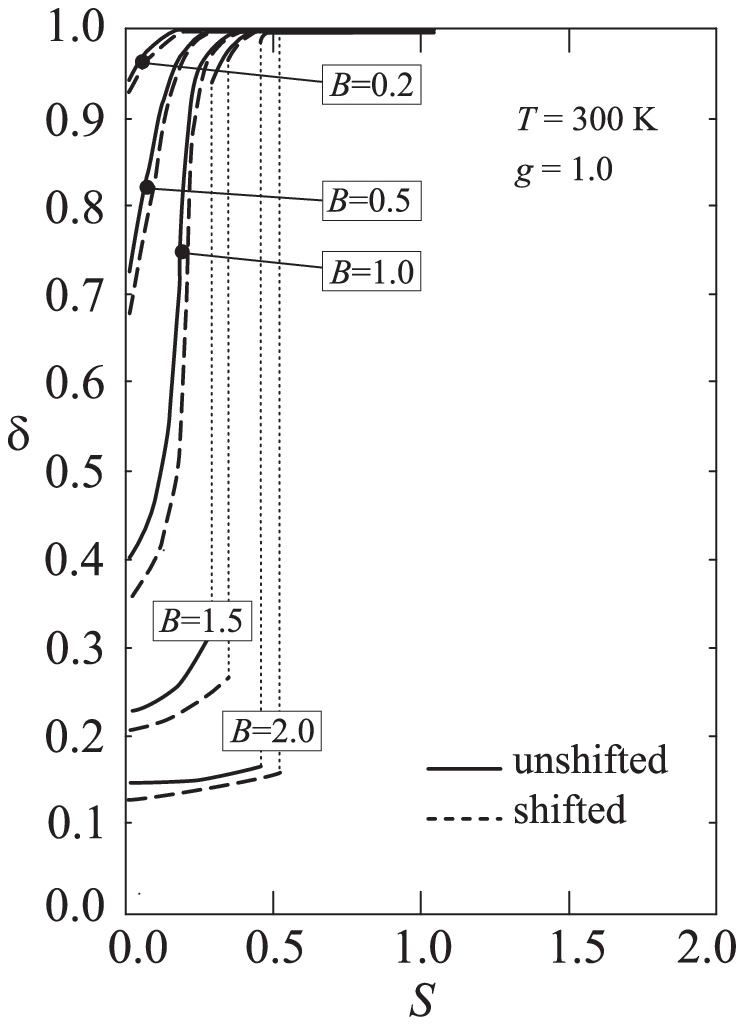}

\includegraphics[height=5 cm]{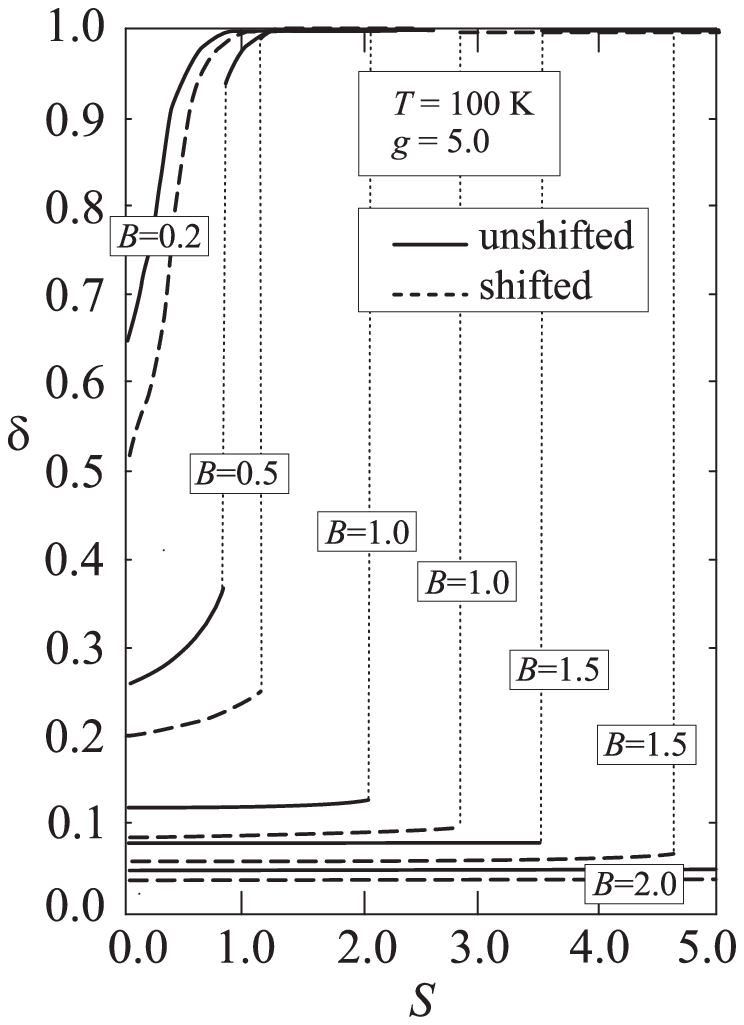}
c
\includegraphics[height=5 cm]{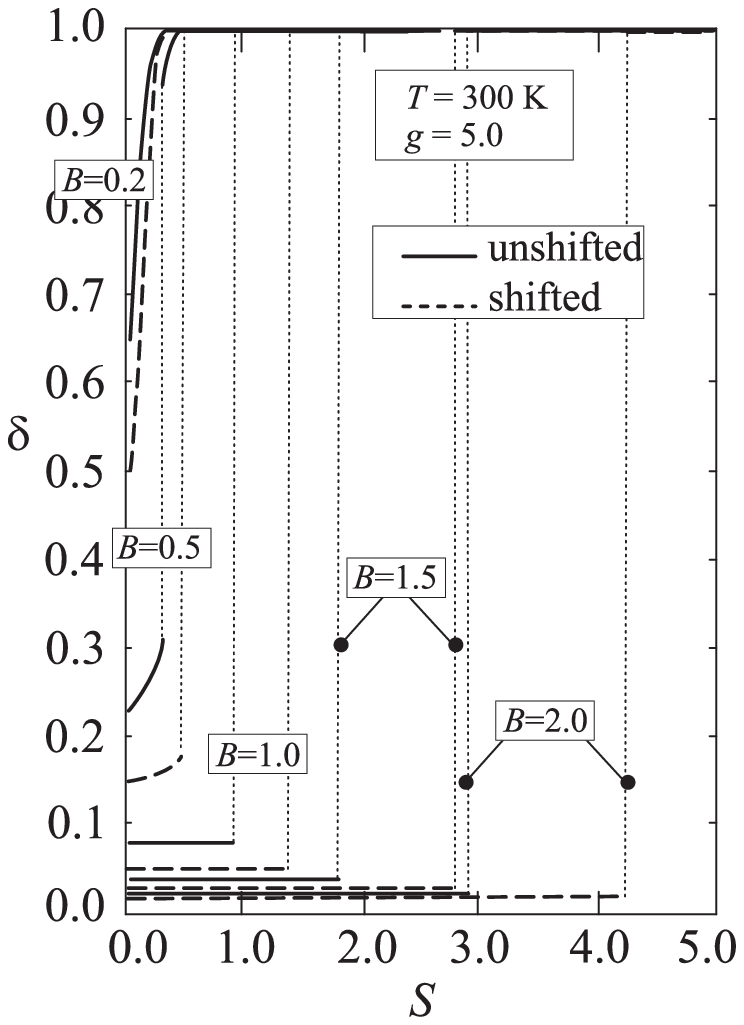}
        \caption{Dressing fraction versus the coupling constant for various values of $B$, $g$, and for two values of system temperature $T$. At each panel we take $T=100$ K (left), and $T=300$ K (right).The values that correspond to unshifted chains are presented by full lines, while those, correspond to shifted MCs are presented by dashed lines. The discontinuous transitions are marked by dotted lines.}\label{f2}
\end{figure}

Optimized dressing fraction as a function of the coupling constant for both geometrical configurations is presented in Fig.2. where we plotted the set of curves $\delta=\delta (S)$ for several values of the adiabatic parameter at two different temperatures and for three different values of the interchain coupling parameter $g$. Values of the adiabatic parameters are chosen to bridge smoothly between the non--adiabatic ($B<<1$) and adiabatic ($B>>1$) limits. It turns out that $\delta$ exhibits the characteristic dependence on coupling constant \cite{BI,PouthierPRB79}: in the non--adiabatic regime it smoothly increases with the coupling constant from some initial value at $S=0$ to unity in the strong coupling limit. As the adiabatic parameter increases, this dependence becomes increasingly sharper: the dressing fraction in the weak and intermediate coupling regime exhibits a smooth and very slow increase with the coupling constant. As $S$ approaches the strong coupling limit, $\delta$ sharply but still continuously approaches unity. When $B$ exceeds the critical value $B_C$, being different for each temperature and for each value of the interchain coupling, this transition becomes discontinuous, so that for each $B>B_C$ an abrupt jump of the $\delta\ll 1$ to $\delta \sim 1$ is observed at the critical value of the coupling constant $S_C=S_C(B)$. The discontinuity in $\delta (S)$ is followed by the sudden jump of the excitation effective mass which leads to a "phase transition" in so called small--polaron cross--over from light free small--polaron band states to the localized heavily dressed, practically immobile excitation. Critical values of the system parameters $S_C$ and $B_C$ for which the vibron suffers the abrupt change of dressing depends on $T$, $g$, and double--chain geometry. We estimated these values numerically, and presented them in (Table \ref{CV}).

\begin{table}[]
\centering
\caption{Critical values of the adiabatic parameter and coupling constant}
\label{CV}
\begin{tabular}{@{}|c||c|c|c|c|@{}}
\hline
      & \multicolumn{2}{c|}{unshifted} & \multicolumn{2}{c|}{shifted} \\ \hline
$(g,T)$ & $S_C$        & $B_C$        & $S_C$         & $B_C$\\ \hline
$(0,100\;\mathrm{K})$        &0.78      &1.8       &0.78         &1.8\\
$(0,300\;\mathrm{K})$        &0.27      &1.8       &0.27         &1.8\\  \hline
$(0.5,100\;\mathrm{K})$      &0.73      &1.6       &0.74         &1.6\\
$(0.5,300\;\mathrm{K})$      &0.25      &1.6       &0.27         &1.6\\  \hline
$(1.0,100\;\mathrm{K})$      &0.67      &1.3       &0.73         &1.3\\
$(1.0,300\;\mathrm{K})$      &0.23      &1.3       &0.24         &1.2\\  \hline
$(5.0,100\;\mathrm{K})$      &0.66      &0.43      &0.68         &0.33\\
$(5.0,300\;\mathrm{K})$      &0.22      &0.42      &0.23         &0.32\\  \hline
\end{tabular}
\end{table}

Such a behaviour is the consequence of the appearance of the double well structure of the system free energy, and these minima may be assigned to a free ($\delta \ll 1$) or localized excitation ($\delta \sim 1$). The values of the system parameters ($S$ and $B$) when the depths of these minima are equal, determine the boundary in the system parameter space ($S-B$--plane) where the transition between the band and localized states occurs (Fig.2).    
 
The impact of temperature is manifested through the reduction of the parameter space where band states may occur. This could be understood on the basis of the fact that, in accordance with $S(\tau)\equiv W(\tau)=S\cal G(\tau)$, temperature effectively increases the magnitude of the coupling constant so that the crossover from free to localized polarons at elevated temperature takes place at lower coupling: temperature lowers the critical adiabatic parameter and critical coupling constants $S_C(B)$ for each adiabatic curve.  

Let us now illustrate the consequences of the variation of the degree of dressing with the system parameters and temperature on vibron dynamics along the macromolecular spine. Note that the coherent polaron motion is restricted to low temperatures, and for that reason it is not relevant for a biological systems \cite{Holstein}. The main contribution to the polaron dynamcis becomes from incoherent polaron hopping, assisted by phonons. Since the Hamiltonian (2) describes dressed vibron with renormalized parameters obtained by MF procedure, it is not appropriate for the calculation of the rate for incoherent vibron motion. This Hamiltonian is used for finding of most favorable vibron states i.e. for optimal value of dressing parameter. For that purpose, we exploit the results of the Holstein pioneering paper \cite{Holstein} adjusted here to account for the effects of the vibron partial dressing:

\begin{equation}
w(\tau)=\frac{\omega_0 B^2}{2\delta}\sqrt{\frac{\pi}{S}\sinh\left(1/2\tau\right)}\mathrm{e}^{-2\delta^2S\tanh \left(1/4\tau\right)},
\end{equation}

\noindent
where the probability of incoherent polaron motion was find from full Hamiltonian, accounting for the residual vibron--phonon interaction within the perturbation theory.

In Fig.3. we plotted the normalized value of $w(S,B)$ for the unshifted configuration for fixed values of $g$ and for two different temperatures. For all values of the adiabatic parameter below the critical value the transition probability smoothly decreases as $S$ increases. When $B$ exceeds the critical "adiabaticity" at particular temperature, $w$ discontinuously drops when $S=S_C(B)$ as a result of the sudden jump of the vibron effective mass. This behaviour simply reflects the discontinuous crossover from SP hopping motion (incoherent, phonon assisted dynamics) towards the immobile SP state.

\begin{figure}[h]
    \begin{center}
\includegraphics[height=5.5 cm]{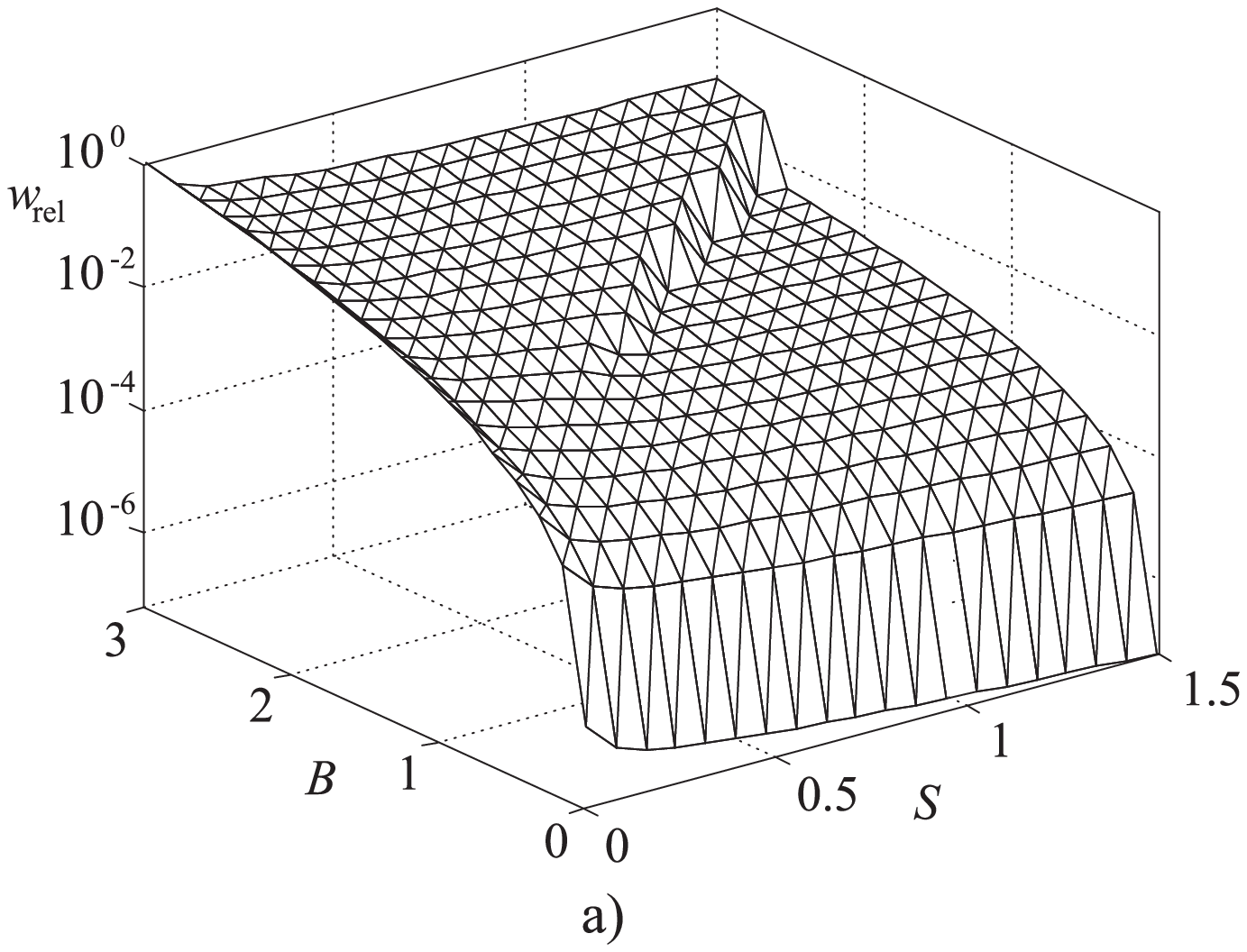}
\includegraphics[height=5.5 cm]{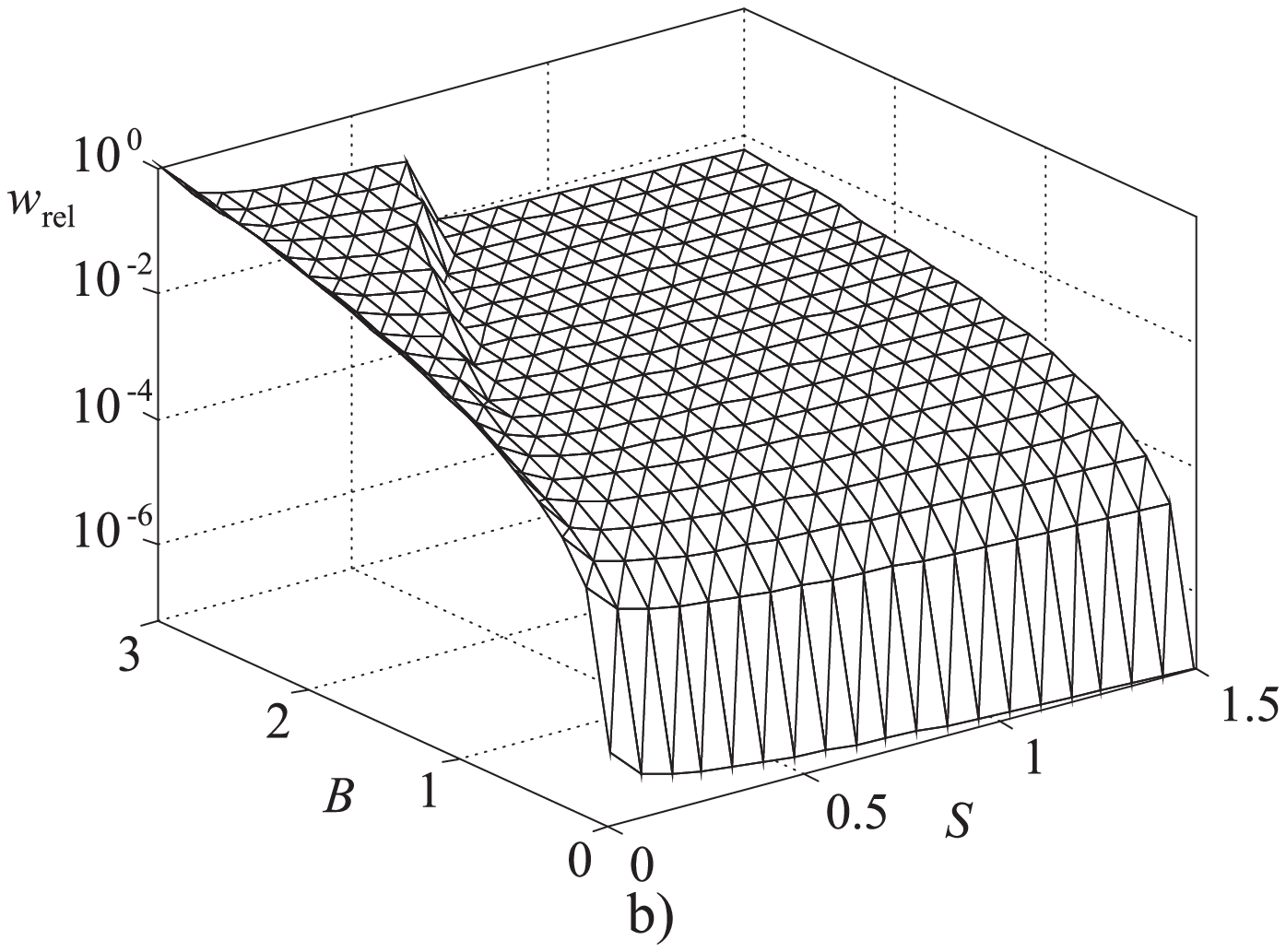}
        \caption{The dependence of the probability of the vibron hopping on parameters $S$ and $B$, for fixed temperature $T=100$ K (graph a)) and $T=300$ K (graph b)). Parameter of anisotropy has the value $g=0.5$.}
    \end{center}
%    \textit{}
\end{figure}

In Fig. 4. we presented the temperature dependence of $w$ for both macromolecular configurations. In the non--adiabatic limit ($B\ll B_C (\tau)$) the probability of vibron hopping is a monotonically decreasing function of normalized temperature $\tau$. As $B$ increases, $w(\tau)$ suffers a discontinuous drop at a particular critical temperature $\tau_C=\tau_C(B)$. For fixed values of $S$ and $g$, $\tau_C$ increases with the increasing of $B$. On the other hand, for fixed value of $B$, $\tau_C$ increases with the increasing of $g$ (as an example, for $g=1.5$ the critical temperature enters to the area of room temperatures, even in the case of unshifted MCs). Similarly to the variational parameter, for small values of $g$ a particular geometric form (shifted or unshifted) has no significant influence on $w$. However, with the increasing of $g$, the difference of the particular values of critical parameters in different structures becomes noticeable. It can also be seen that $w(\mathrm{shifted})>w(\mathrm{unshifted})$ for intermediate and room temperatures.    

\begin{figure}[h]
    \begin{center}
\includegraphics[height=5.5 cm]{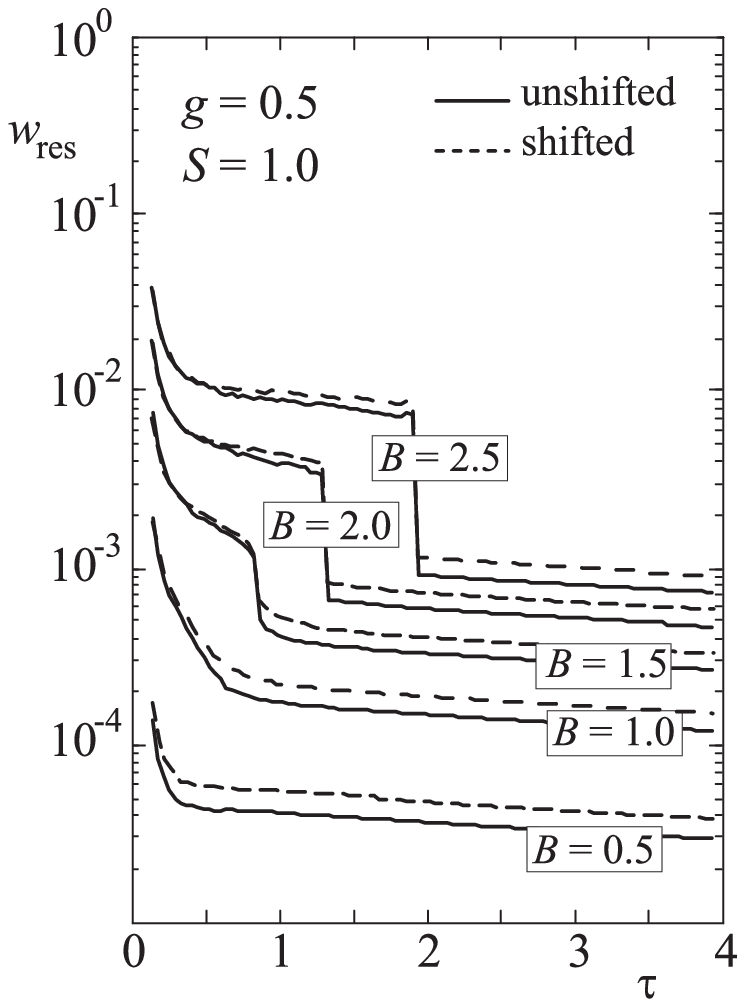}
\includegraphics[height=5.5 cm]{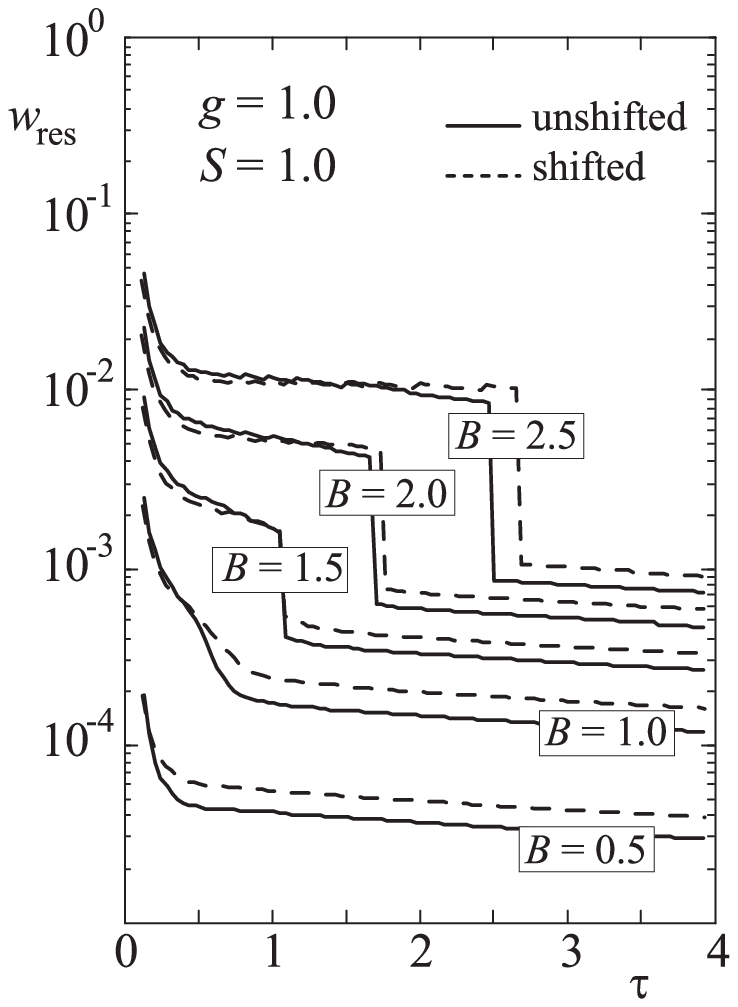}
\includegraphics[height=5.5 cm]{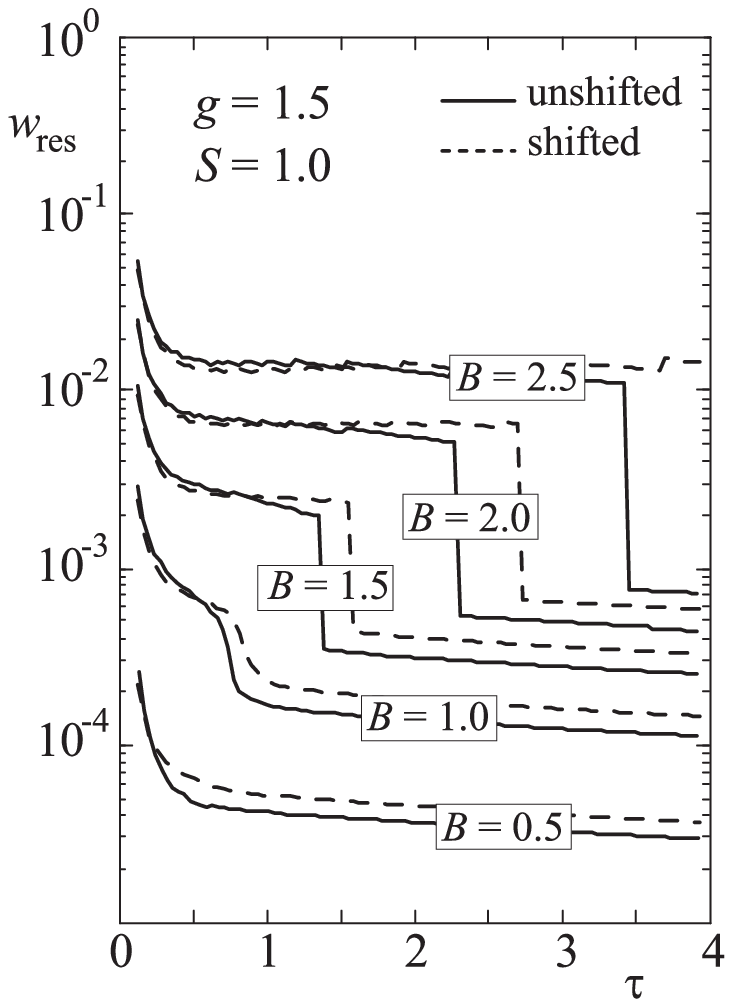}
        \caption{The dependence of the probability of the vibron hopping on normalized temperature $\tau$ for various values of $B$ and $g$, and for both MCs configuration.}
    \end{center}
%    \textit{}
\end{figure}

From the presented results we can conclude that ST in two chain media exhibits qualitatively the same features as in 1D ones, the only difference is in the value of the critical parameters which are lower than in the pure 1D case. Our results suggest that for large interchain coupling the phase transition can get into the non--adiabatic and weak interaction region of system parameters. This especially concerns the high temperature region. The coupling facilitates migration, but for anti--adiabatic limit at finite temperatures it is still not enough to produce the coherent transfer. Just like in 1D model the vibron motion in coupled chains has the incoherent character: it takes place in a view of random jumps between neighbouring sites.

Due to the fact that results presented here correspond to the case when IMO interacts with nondispersional optical phonons, it may be interesting to investigate the case of the IMO interaction with acoustic phonon modes. For that case, vibron transfer between different chains becomes temperature dependent and the picture of the vibron dynamics can be changed.

Obtained results can be useful for the better understanding of the process of the energy transport in such structures that are consist of two (or more) coupled macromolecular chains, like DNA macromolecule, for example. In spite of all efforts to understand such process, this problem still remains poorly understood \cite{Zaoli_AIPadv4}. At the present, it is known that DNA is poor heat conductor wit a low thermal conductivity \cite{VelizhaninPRE83,SavinPRB83}, but its thermal properties may significantly alter in dependence of preparation of DNA composited samples \cite{VelizhaninPRE83}. For that reasons, we believe that it may be interesting to investigate the possible role of the partially dressed polarons (including partially dressed vibrons) in the process of energy and charge transfer along the DNA spine.

In conclusion we discuss the reliability of our results. We first note that influence of temperature and system parameters on interchain vibron transfer cannot be seen from our analysis. This is the consequence of the fact that the present model accounts for vibron interaction with non-- dispersive optical phonons for which we found the equal dressing degree for longitudinal and transverse effective mass: $m_l/m_t=J_{eff}/L_{eff}$, where $J_{eff}=J\mathrm{e}^{-\delta^2W(\tau)}$, and $L_{eff}=L\mathrm{e}^{-\delta^2W(\tau)}$. This imply $m_l=g\cdot m_t$. As a consequence, for $g<1$ we have $m_l<m_t$, which favors the vibron localization to the single chain. When $g>1$ we have that $m_l>m_t$ which favors the vibron tunneling among the chains. In that respect, our results differs from those presented in \cite{FalvoJCP123a} devoted to single vibron state in $\alpha$-- helix macromolecules, where polaron exchange among different chains strongly depends on temperature. In low temperature (or weak coupling regime) vibron is weakly dressed while its dynamics takes place as random jumps between different spines. With the increase of the temperature (or with the increasing of the coupling) the dressing effect strongly reduces the vibron transfer between different chains, and it migrates along a single spine.

We finally we must point out that the aforementioned numerical confirmation of the discontinuous small polaron crossover does not mean full "rehabilitation" of the TME and related variational theories. In particular, it may be regarded as satisfactory in the nonadiabatic regime, while in the adiabatic regime the consistency with the exact numerics is merely qualitative since their predictions underestimates the values of critical parameters. For that purpose, application of the present results in the particular context would be taken with certain caution.

\begin{center}
\textbf{Acknowledgments}
\end{center}

This work was supported by the Serbian Ministry of Education and Science, under Contract Nos. III-45010, III-45005, OI--171009, and by the Project within the Cooperation Agreement between the JINR, Dubna, Russian Federation and Ministry of Education and Science of Republic of Serbia.

\end{document}